\documentclass[a4paper,12pt]{article}
\usepackage[utf8x]{inputenc}
\usepackage{amssymb,amsfonts,amstext,amsmath,graphicx,epic,amsthm,color,times}

\title{The inviscid instability in an electrically conducting fluid affected by a parallel magnetic field}
\author{A. V. Monwanou\footnote{movins2008@yahoo.fr} and J. B. Chabi Orou\footnote{Author to whom correspondence should be addressed: jchabi@yahoo.fr}}
\date{}
\begin{document}

\maketitle Institut de Math\'ematiques et de Sciences Physiques, BP: 613 Porto Novo, B\'enin

The Abdus Salam International Centre for Theoretical Physics, Trieste, Italy

\begin{abstract}
We investigate inviscid instability in an electrically conducting fluid affected by a parallel magnetic field. The case of low magnetic
 Reynolds number in Poiseuille flow is considered. When the magnetic field is sufficiently strong, for a flow with low hydrodynamic 
Reynolds number, it is already known that the neutral disturbances are three-dimensional. Our investigation shows that at 
high hydrodynamic Reynolds number(inviscid flow), the effect of the strength of the magnetic field on the fastest growing
 perturbations is limited to a decrease of their oblique angle i.e. angle between the direction of the wave propagation and the basic
 flow. The waveform remains unchanged. The detailed analysis of the linear instability provided by the eigenvalue problem shows that 
the magnetic field has a stabilizing effect on the electrically conducting fluid flow. We find also that at least, the unstability appears if the main flow possesses an inflexion
 point with a suitable condition between the velocity of the basic flow and the complex stability parameter according
 to Rayleigh's inflexion point theorem. 
\end{abstract}

\section{Introduction}
 We consider the instability in a shear flow of an incompressible viscous electrically conducting fluid with the initial 
velocity profile\textbf{\cite{Hea}}
\begin{equation}
 \textbf{U}=(U(z),0,0),\;\; U(z)\rightarrow 0\;\; at\;\; z=(z_1, z_2)=(-1, +1).
\end{equation}
Next, we impose throughout the flow a uniform time-independent magnetic field $\textbf{B}=(B,0,0)$ in the streamwise direction. 
The magnetic Reynolds number will be assumed to be small\textbf{\cite{Nar}} i.e.
\begin{equation}
 Re_m=\frac{U_0 L}{\lambda}<<1,
\end{equation}
where $L$ is the length scale and will be taken as the initial vorticity thickness of the layer, $U_0$ represents the velocity scale for
 the flow and $\lambda=\frac{1}{\sigma \mu_0}$ stands for the magnetic diffusivity in which $\sigma$ is the electrical conductivity of the
 fluid and $\mu_0$ the magnetic permeability of a vacuum. 

The condition given by formula $(2)$ is widely obtained in industrial flows or liquid metals, molten oxides etc... This allows one to
 apply the low-$Re_m$ approximation (Davidson 2001) in which only the imposed magnetic field \textbf{B} in the Lorentz force expression 
is taken into account. This leads to the following non-dimensional equations
 \begin{eqnarray}
   \begin{array}{c}
 \frac{\partial \textbf{v}}{\partial t}+(\textbf{v}.\nabla)\textbf{v}=-\nabla p +
\frac{1}{Re}\nabla^2 \textbf{v}+N(\textbf{j}\wedge\textbf{B}),\\\\
\nabla.\textbf{v}=0.
\end{array} 
\end{eqnarray}

 The electric current \textbf{j} is given by\textbf{\cite{Bhi}} 
\begin{equation}
 \textbf{j}=-\nabla\phi+\textbf{v}\wedge\textbf{B},
\end{equation}
where $\phi$ is the electric potential which is a solution of the Poisson equation  
\begin{equation}
\nabla^2 \phi=\textbf{B}.(\nabla\wedge\textbf{v}).
\end{equation}

The two non-dimensional parameters appearing in eq.3 are defined as 
\begin{equation}
 Re=\frac{U_0 L}{\nu}, \;\; N=\frac{\sigma B^2 L}{\rho U_0}
\end{equation}
(Reynolds number and magnetic interaction parameter respectively).

The magnitude of $N$ gives information on the ratio between the Lorentz and inertia forces leading to the evaluation of the
 potential of the magnetic field which suppresses and transformes the perturbations.

There are no electric or Lorentz forces generated in the non-perturbed basic flow \textbf{\cite{Vor}}.
 
\section{Governing equations}
By using the stability analysis of a shear velocity profile in the presence of a parallel magnetic field performed by Michael (1953),
 Stuart (1954) and Drazin (1960), we can use the normal modes for the fluctuating part of the velocity in the form

\begin{equation}
 \textbf{v}^{'}(x,y,z,t)=\textbf{v}(z)exp[i(k_x x+k_y y-\lambda t)]
\end{equation}

in the standard way of linear stability analysis with  $k_x$ and $k_y$ representing the real wavenumbers in the x- and y-directions,
 $\lambda=\omega +i \beta$ is the complex stability parameter, where $\beta$ is the growth rate of the instability and $\omega$ is 
the frequency. If $\beta > 0$ , the disturbance grows and the system becomes unstable. Whereas, if $\beta<0$, the disturbance decays
and the system becomes stable. $\beta=0$ corresponds to neutral instability.

At this point, one should point out that the magnetic field stabilizes the flow because of the Joule dissipation action which 
suppresses the growing perturbations \textbf{\cite{Vor}}. But it has to be verified by an eigenvalue problem
 where $\lambda=\omega +i \beta$ will be the eigenvalue.

We consider here only two-dimensional disturbances \textbf{\cite{Lan}}  with $k_y=0$. For an arbitrary mode with $k_y\neq 0$, 
the classical generalized Orr-Sommerfeld equation becomes
\begin{equation}
 (k_x U-\lambda)(v_z^{''}-k^2 v_z)-k_x v_z U^{''}+ik_x^2 N v_z=-\frac{i}{Re}(v_z^{''''}-2 k^2 v_z^{''}+k^4 v_z), 
\end{equation}
for which $v_z=v'_z=0$ if $z=(z_1, z_2)$.

As usual,  $k=(k_x^2+ k_y^2)^{\frac{1}{2}}$, the primes
 stand for the first derivatives with respect to z. $\theta=cos^{-1}(\frac{k_x}{k})$ is called the oblique angle 
between the direction of the wave propagation and the basic flow.

By rearranging $(8)$, we obtain 
\begin{equation}
 (U-\tilde{\lambda})(v_z^{''}-k^2 v_z)-v_z U^{''}+ik \tilde{N} v_z=-\frac{i}{k \tilde{Re}}(v_z^{''''}-2 k^2 v_z^{''}+k^4 v_z), 
\end{equation}
with the following boundary conditions:  $v_z=v'_z=0$ if $z=(z_1, z_2)$.

We redefine new non-dimensional parameters as follow: $\tilde{Re}=(\frac{k_x}{k})Re$, $\tilde{N}=(\frac{k_x}{k})N$ and 
$\tilde{\lambda}=(\frac{\lambda}{k_x})$.
 
The solution of the problem is given as a relation between $\tilde{\lambda}$,  $\tilde{Re}$, $\tilde{N}$ and $k$ in the form 
\begin{equation}
 F(\tilde{\lambda},k,\tilde{Re},\tilde{N})=0, 
\end{equation}
for any angle $\theta$. A particular solution can be determined for a two-dimensional waveforms with $\theta=0$. We could get a 
critical Reynolds number $Re_c$ which corresponds to the minimum $Re$ occurring over all $k$ and $\omega$ at which a neutral mode with 
$\beta=0$ is noticed by writing
\begin{equation}
 F_c(\tilde{Re_c},\tilde{N})=0 \;\; or \;\; \tilde{Re_c}=G(\tilde{N}).
\end{equation}
In the non-magnetic case with $N=0$, the Squire theorem \textbf{\cite{Squ}, \cite{Lan}} requires that the two-dimensional perturbations are always 
first to become unstable since the smallest critical Reynolds number $Re_c=\frac{k}{k_x}\tilde{Re_c}=\frac{\tilde{Re_c}}{cos\theta}$ 
is always for the perturbations with $\theta=0$.

For the inviscid flow, we have 
\begin{equation}
 Re\rightarrow \infty, 
\end{equation}
and then the generalized Orr-Sommerfeld problem (9) becomes 
\begin{equation}
  (k_x U-\lambda)(v_z^{''}-k^2 v_z)-k_x v_z U^{''}+ik_x^2 N v_z=0
\end{equation}
with the condition $v_z=v'_z=0$ if $z=(z_1, z_2)$.

By doing as $(9)$, we have
\begin{equation}
 (U-\tilde{\lambda})(v_z^{''}-k^2 v_z)-v_z U^{''}+ik \tilde{N} v_z=0
\end{equation}
with the condition $v_z=v'_z=0$ if $z=(z_1, z_2)$.

Here, the non-dimensional parameters are $\tilde{N}=(\frac{k_x}{k})N$, and $\tilde{\lambda}=(\frac{\lambda}{k_x})$.

\section{Rayleigh's inflexion point for the flow}
Let us consider the linear stability of a uni-directional base flow in a channel. We derive the Orr-Sommerfeld equation, which 
governs the linear stability of uni-directional shear flows with respect to 3D perturbations, for viscous fluids. We obtain it by 
taking $N=0$ in (9).Then we can write
\begin{equation}
 (U-\tilde{\lambda})(v_z^{''}-k^2 v_z)-v_z U^{''}=-\frac{i}{k \tilde{Re}}(v_z^{''''}-2 k^2 v_z^{''}+k^4 v_z), 
\end{equation}
with the condition $v_z=v'_z=0$ if $z=(z_1,z_2)$.

In the inviscid case, we have the following Rayleigh's equation
\begin{equation}
 (U-\tilde{\lambda})(v_z^{''}-k^2 v_z)-v_z U^{''}=0. 
\end{equation}
Suppose that $U$ and $DU$ where $D=\frac{d}{dz}$ are continuous in $z_1<z<z_2$. Rayleigh's inflexion point theorem then states that
 a necessary (though
 not sufficient) condition for inviscid instability is that the base state possesses an inflexion point somewhere in the 
domain $z_1<z<z_2$. If a base state lacks an inflexion point, therefore, we can conclude it to be stable, for inviscids fluids.

 Consider equation (16) in the following form with the substitution $v_z=v$, 
\begin{equation}
 D^2 v-(k^2+\frac{D^2 U}{U -\tilde{\lambda}})v=0.
\end{equation}
Suppose initially that the flow is unstable ($\beta>0$), it is proved that an inflexion point i.e $D^2 U=0$ must exist for this to be so.

Using boundary condition $v(z_1)=v(z_2)=0$ and by making some calculations, we get
\begin{equation}
 -\int^{z_2}_{z_1} |Dv|^2 dz-\int^{z_2}_{z_1} (k^2+\frac{D^2 U .(U-\bar{\tilde{\lambda}})}{|U-\tilde{\lambda}|^2})|v|^2 dy=0
\end{equation}
where $\bar{\tilde{\lambda}}$ is the complex conjugate of $\tilde{\lambda}$.
The imaginary part of this equation is 
\begin{equation}\label{im}
 -\beta \int^{z_2}_{z_1} \frac{D^2 U |v|^2}{|U-\tilde{\lambda}|^2} dy=0.
\end{equation}
From the hypothesis $\beta>0$, we conclude that $D^2 U$ must change signe somewhere in the domain $(z_1,z_2)$.

Then, a necessary condition for inviscid instability is the presence of an inflexion point; the absence of an inflexion 
point necessarily confers (inviscid) stability.

Let us investigate what happens if the flow is affected by a parallel magnetic field i.e the case $N \neq 0$.

The same calculation leads to
\begin{equation}
 -\int^{z_2}_{z_1} |Dv|^2 dz-\int^{z_2}_{z_1} \left[k^2+\frac{(D^2U-ik\tilde{N})(U-\bar{\tilde{\lambda}})}{|U-\tilde{\lambda}|^2}\right]|v|^2 dy=0.
\end{equation}

The imaginary part of this equation is
\begin{equation}\label{ima}
 - \int^{z_2}_{z_1} \frac{ \beta D^2 U+k \tilde{N} (\tilde{\omega}-U)}{|U-\tilde{\lambda}|^2}|v|^2 dy=0.
\end{equation}
 (\ref{im}) and (\ref{ima}) lead to same properties if 

\begin{equation}
k \tilde{N} (\tilde{\omega}-U)=0
\end{equation}
with $\tilde{\lambda}=\tilde{\omega}+i\tilde{\beta}.$

So the inviscid flow will be unstable if there is the presence of an inflexion point in the main flow with the condition
\begin{equation}
 \tilde{\omega}=U(z)\;\;\; i.e\;\;\; \Re_e(\tilde{\lambda})=U(z).
\end{equation}

\section{Linear stability analysis}
We analyse linear stability of the basic flow (1) to normal mode (7). A Poiseuille flow with the basic profile 
\begin{equation}
 U(z)=1-z^2
\end{equation}
is considered.

The eigenvalue problem (14) is solved numerically. The solution is found in a layer bounded at $z=\pm 1$ with $U(\pm1)=0$. 
The results of calculations are presented in the following figures. 

For a fixed $\theta=0$, we get figure 1 of $\beta$ vs k in which $a)$ shows the entire graph. $b)$ and $c)$ are the magnified versions 
of $a)$.

For sequential values ​​of $N$, we get figures 2, 3 and 4 of $\beta$ vs k for differents $\theta$ in which a) shows 
the entire graph. b) and c) are the magnified versions of a).

We get also figure 5 of $\tilde{\beta}$ vs k for differents $\tilde{N}$ in which $a)$ shows 
the entire graph. $b)$ and $c)$ are the magnified versions of $a)$.

For sequential values ​​of $N$, we get figures 6, 7 and 8 of $\tilde{\beta}$ vs k for differents $\theta$ in which $a)$ shows 
the entire graph. $b)$ and $c)$ are the magnified versions of $a)$.\\\\
\begin{figure}[htbp]
  \begin{center}
  \includegraphics[width=9cm]{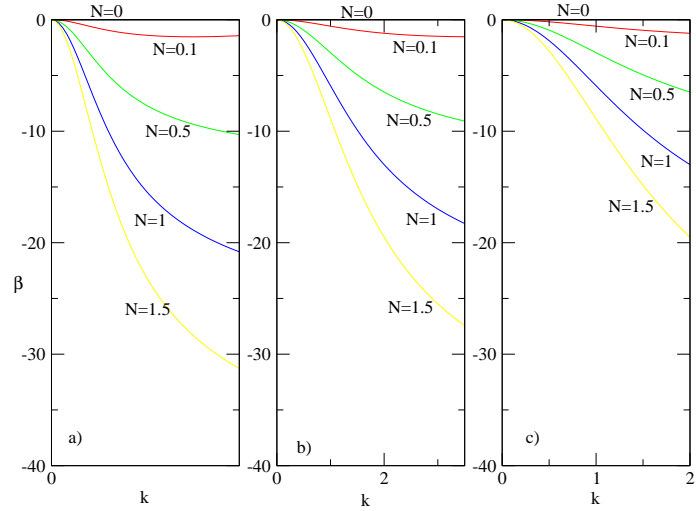}
  \end{center}
  \caption{(a) Growth rate $\beta$ vs. wavenumber k with $\theta=0$; (b) zoom of (a) to small values of k; (c) zoom of (b) to small
 values of k. }
 \label{yghl}
 \end{figure}

\begin{figure}[htbp]
  \begin{center}
  \includegraphics[width=10cm]{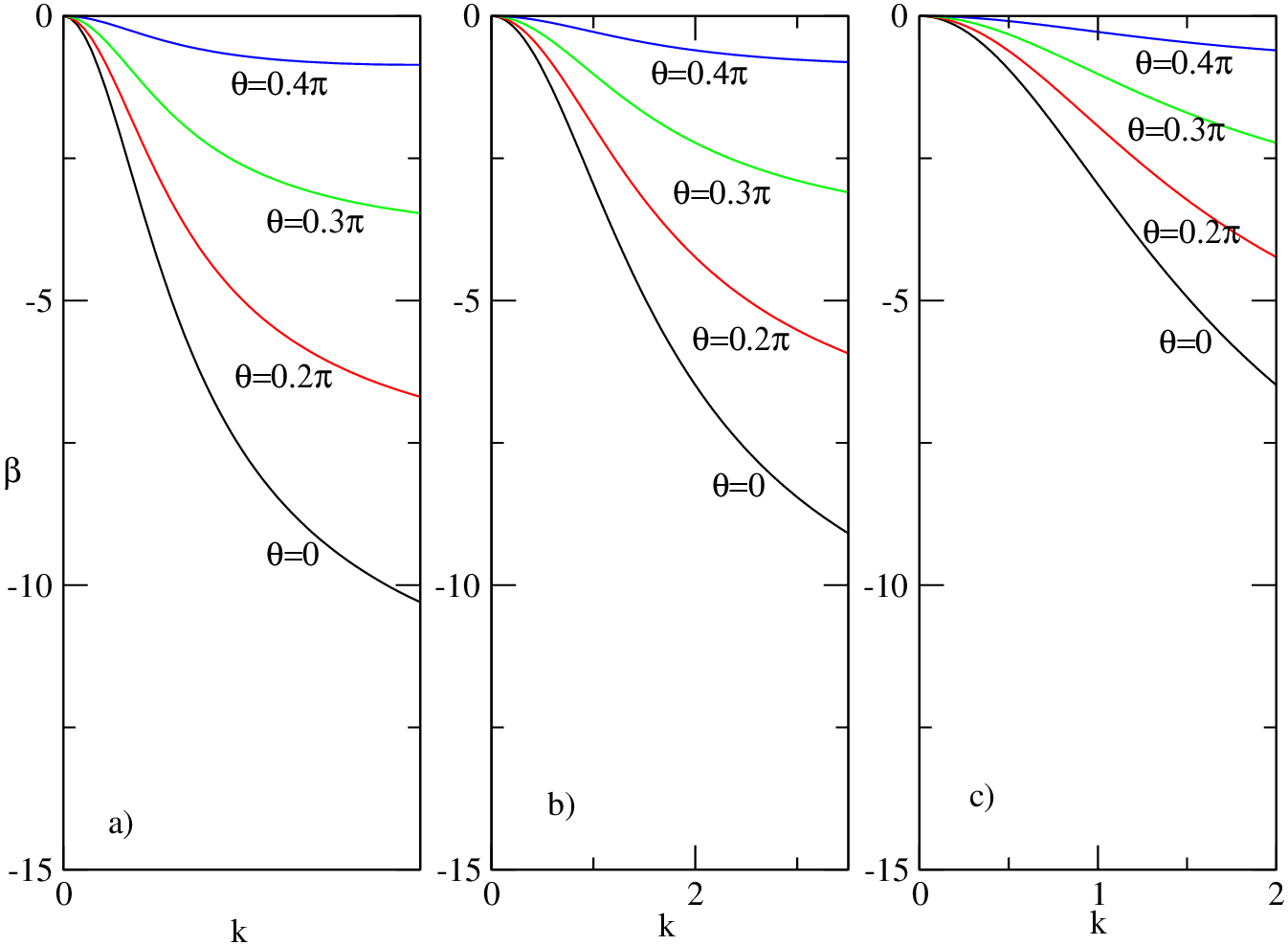}
  \end{center}
 \caption{(a) Growth rate $\beta$ vs. wavenumber k. with $N=0.5$; (b) zoom of (a) to small values of k; (c) zoom of (b) to small
 values of k. }
 \end{figure}

\begin{figure}[htbp]
  \begin{center}
  \includegraphics[width=10cm]{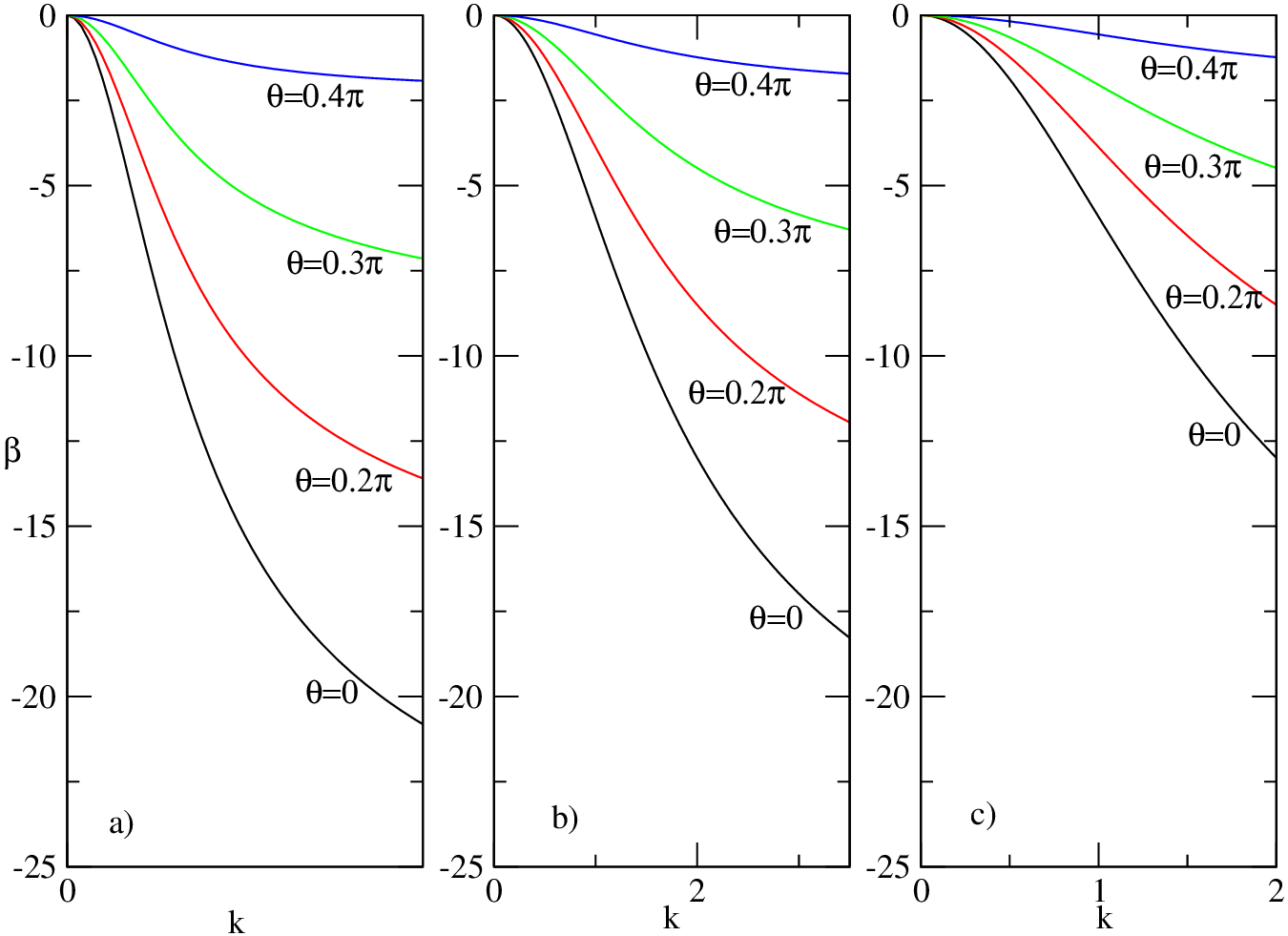}
  \end{center}
 \caption{(a) Growth rate $\beta$ vs. wavenumber k with $N=1$; (b) zoom of (a) to small values of k; (c) zoom of (b) to small
 values of k.}
 \end{figure}

\begin{figure}[htbp]
  \begin{center}
  \includegraphics[width=10cm]{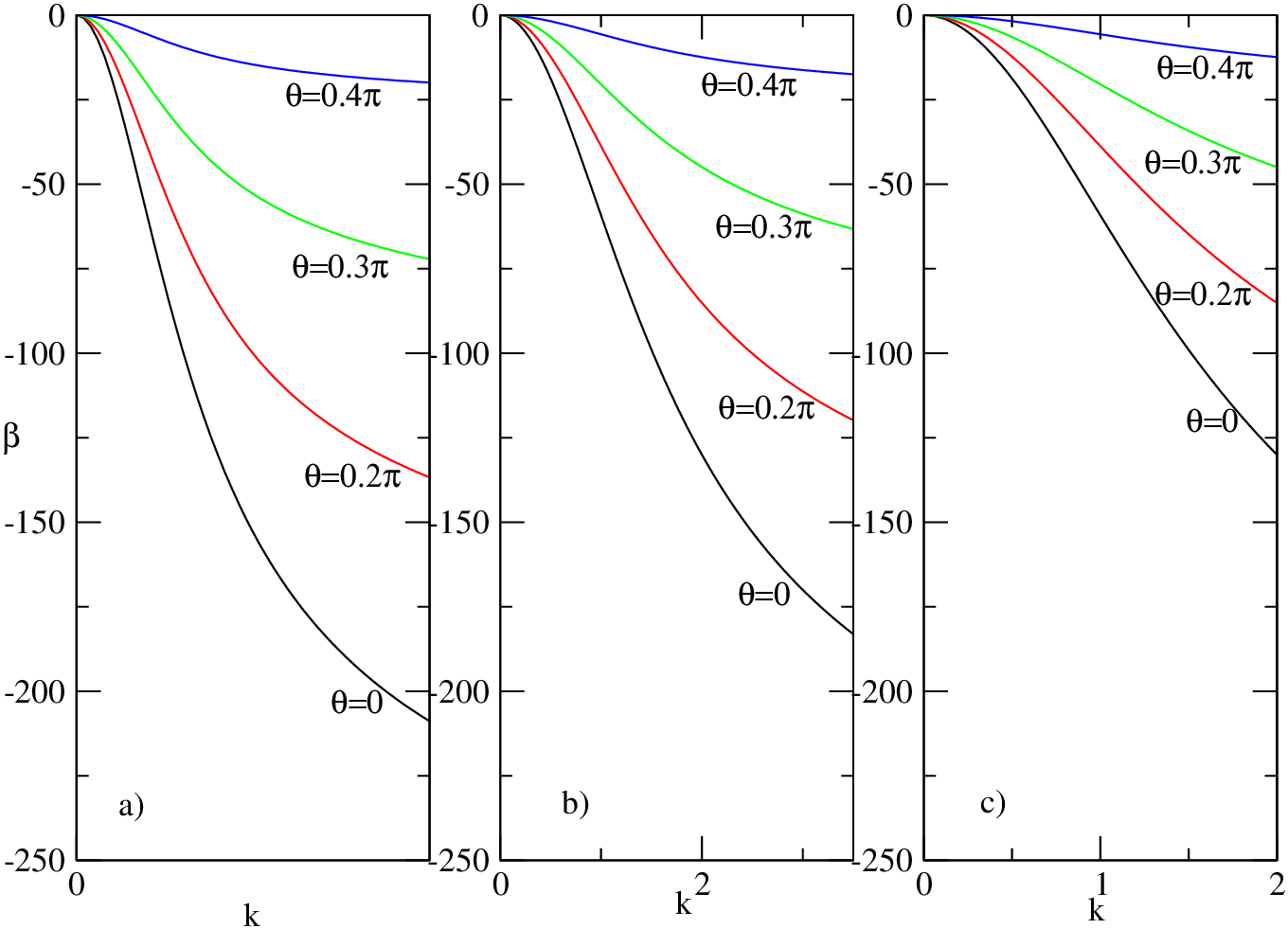}
  \end{center}
 \caption{(a) Growth rate $\beta$ vs. wavenumber k with $N=10$; (b) zoom of (a) to small values of k; (c) zoom of (b) to small
 values of k.}
 \end{figure}

\begin{figure}[htbp]
  \begin{center}
  \includegraphics[width=10cm]{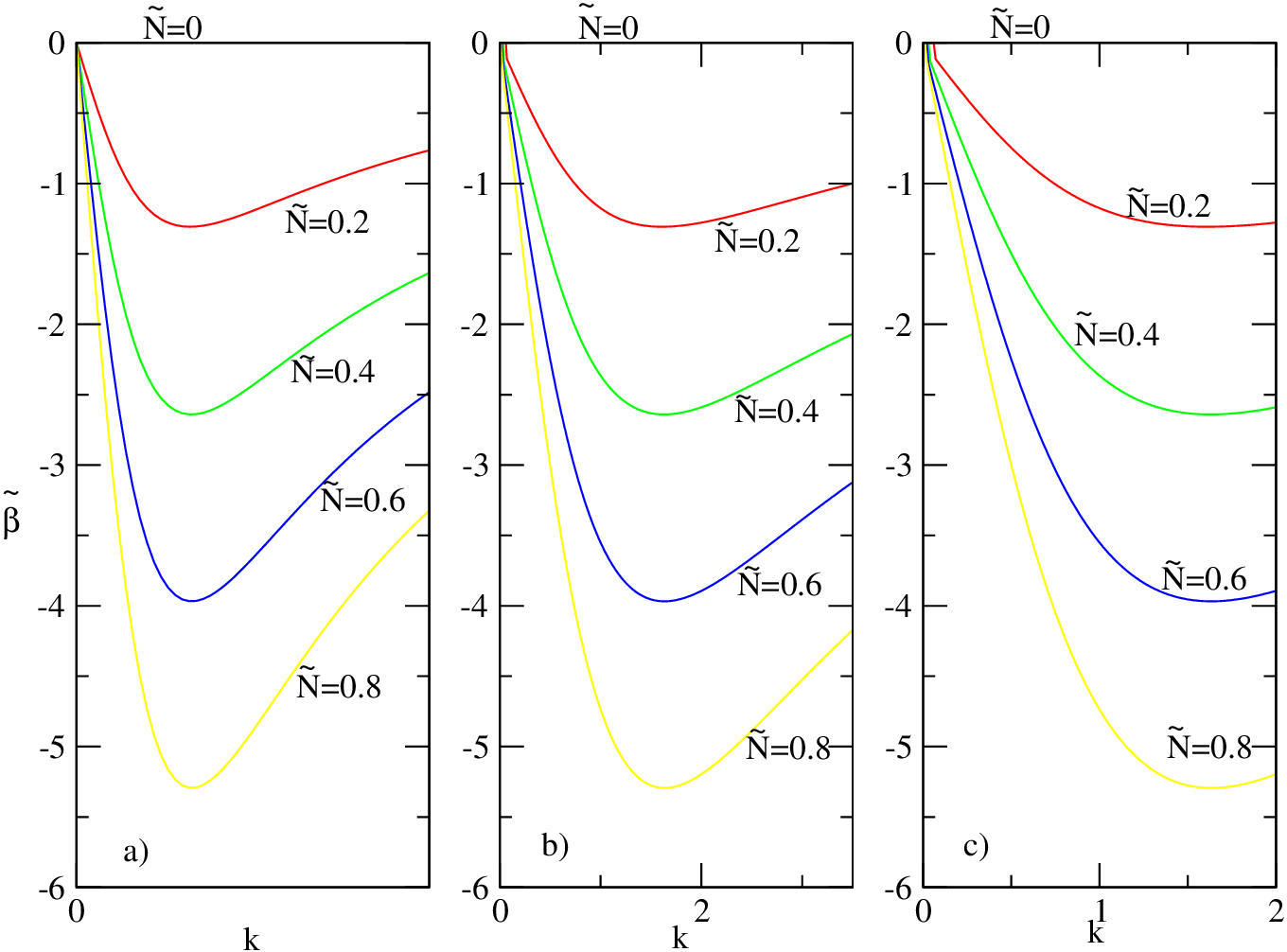}
  \end{center}
  \caption{(a) Growth rate $\tilde{\beta}$ vs. wavenumber k ; (b) zoom of (a) to small values of k; (c) zoom of (b) to small
 values of k.}
 \end{figure}

\begin{figure}[htbp]
  \begin{center}
  \includegraphics[width=10cm]{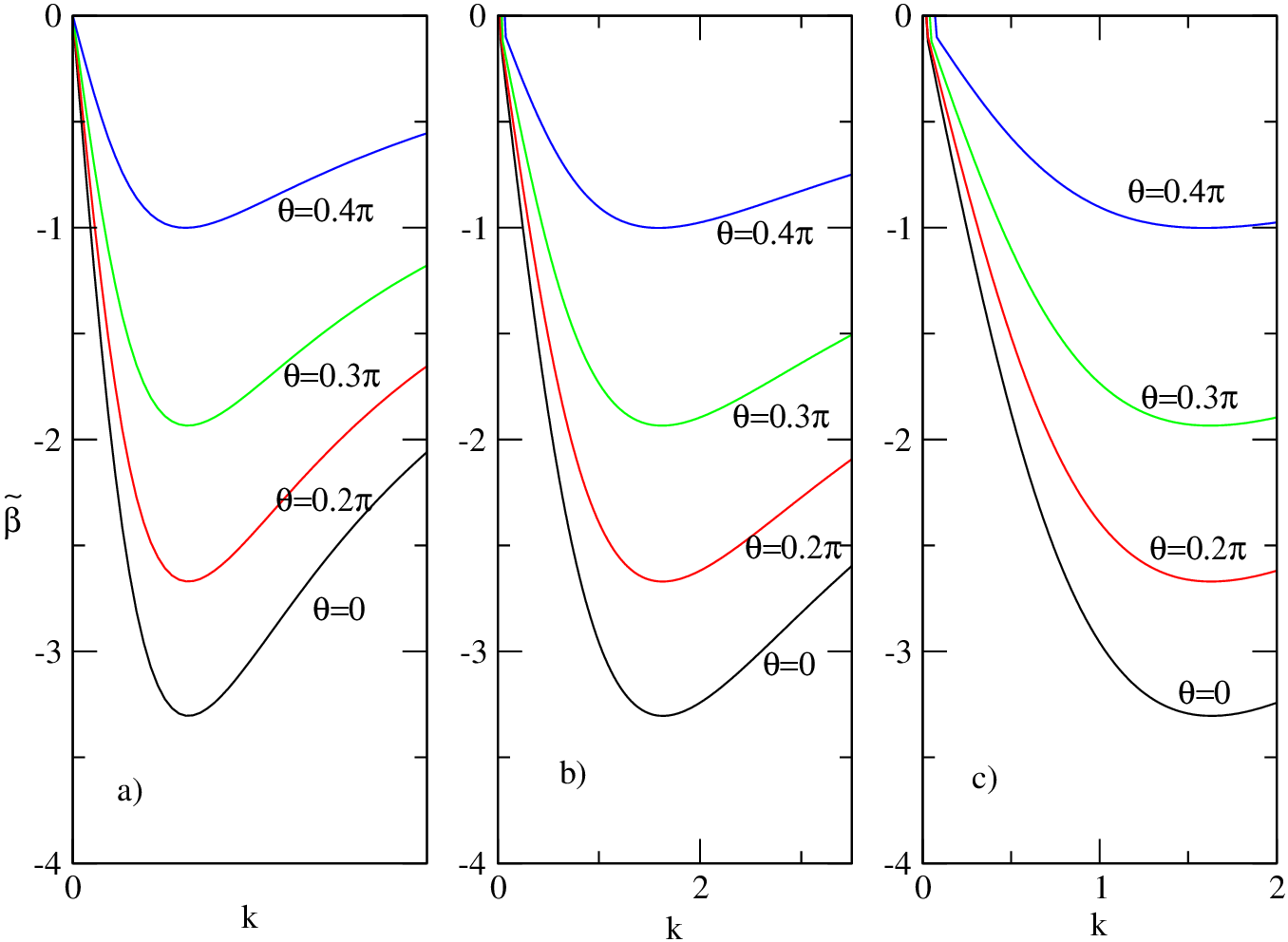}
  \end{center}
 \caption{(a) Growth rate $\tilde{\beta}$ vs. wavenumber k with $N=0.5$; (b) zoom of (a) to small values of k; (c) zoom of (b) to small
 values of k.}
 \end{figure}

\begin{figure}[htbp]
  \begin{center}
  \includegraphics[width=10cm]{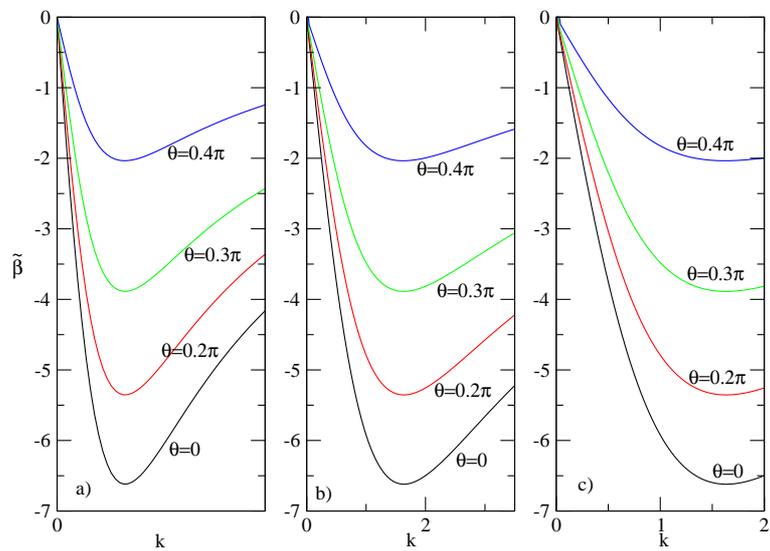}
  \end{center}
\caption{(a) Growth rate $\tilde{\beta}$ vs. wavenumber k with $N=1$; (b) zoom of (a) to small values of k; (c) zoom of (b) to small
 values of k.}
 \end{figure}

\begin{figure}[htbp]
  \begin{center}
  \includegraphics[width=10cm]{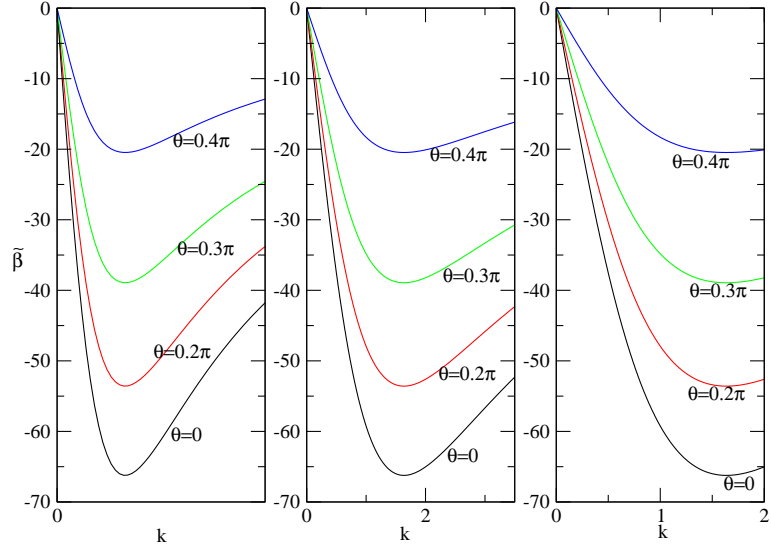}
  \end{center}
\caption{(a) Growth rate $\tilde{\beta}$ vs. wavenumber k with $N=10$; (b) zoom of (a) to small values of k; (c) zoom of (b) to small
 values of k.}
 \end{figure}
\newpage
Figures 1-8 show the strong stabilizing effect of the magnetic field on the two-dimensional perturbations.

It has to be stressed that the complete stabilization requires non-zero viscosity. It is shown in the inviscid two-dimensional 
analysis of Thess \& Zikanov (2005) that the shear flow (1) cannot be completely stabilized by the magnetic field. There 
always exists a range of small $k$, where the flow is unstable. Such behaviour is in agreement with the intuitive pictures, 
according to which the rate of the Joule dissipation decreases with increasing wavelength in the direction of the magnetic 
field, and, thus, the perturbations become less and less sensitive to the action of the magnetic field as $k\rightarrow 0$.

Typical dependence of $\beta$ on $\theta$ and $k$ for the three-dimensional disturbances is shown in figures 2-4. The growth 
rate changes slowly with the wavenumber and the oblique angle.

\section{Conclusion}
In this paper, we revisited the inviscid instability of an electrically conducting fluid(modelled as a temporally evolving flow
 initially given 
by a Poiseuille flow velocity profile) subject to a parallel uniform magnetic field. The case of small magnetic Reynolds number 
was considered. We find an important condition between the velocity of the basic flow and the complex stability parameter for which 
the main flow, if it possesses an inflexion point, leads to unstability according to Rayleigh's inflexion point theorem. We provided 
detailed analysis
 of the linear instability of the problem in Poiseuille case in direct numerical simulations by resolving the corresponding eigenvalue 
problem. It shows us that the magnetic field has a stabilizing effect on the electrically conducting fluid; however, it remains stable
 for all possible
 values of the magnetic field since the wavenumber is non-zero.

\section*{Acknowledgments}
The authors thank IMSP-UAC for financial support.

\end{document}